\documentclass[sn-mathphys]{sn-jnl}

\usepackage{graphicx}%
\usepackage{multirow}%
\usepackage{amsmath,amssymb,amsfonts}%
\usepackage{amsthm}%
\usepackage{mathrsfs}%
\usepackage[title]{appendix}%
\usepackage{xcolor}%
\usepackage{textcomp}%
\usepackage{manyfoot}%
\usepackage{booktabs}%
\usepackage{algorithm}%
\usepackage{algorithmicx}%
\usepackage{algpseudocode}%
\usepackage{listings}%

\usepackage{times}
\usepackage{graphics}
\usepackage{amsmath}
\usepackage{amsfonts}
\usepackage{amssymb}
\usepackage{subfigure}
\usepackage{xcolor}
\usepackage{multirow}
\usepackage{bigints}
\usepackage{mathtools}
\usepackage{bm}

\usepackage{natbib}
\usepackage{verbatim}
\usepackage{graphicx}
\usepackage{psfrag}
\usepackage{bm}
\usepackage{array}
\usepackage{amsmath}
\usepackage{amssymb}
\usepackage{xcolor}
\usepackage{makeidx}         
\usepackage{graphicx}        
\usepackage[bottom]{footmisc}
\usepackage{natbib}		
\usepackage{multirow}
\usepackage{subfigure}

\theoremstyle{thmstyleone}%
\newtheorem{theoremcompliant}{Proposition}

\def\Ma{\mbox{Ma}}

\def\Masq{\mbox{Ma}^2}

\def\bu{{\bf u}}
\def\disp{\eta}

\def\co{c_o}

\def\baru{\bar{u}}

\def\barrho{\bar{\rho}}

\def\barT{\bar{T}}

\def\barp{\bar{p}}

\def\maxbaru{\mbox{Max}({\bar{u}})}
\def\minbaru{\mbox{Min}({\bar{u}})}

\def\kx{k_x}
\def\kz{k_z}
\def\kxsq{k_x^2}
\def\kzsq{k_z^2}

\def\cR{c_R}
\def\cI{c_I}
\def\yc{y_c}
\def\ca{c^\ast}

\def\tg{\hat{g}}
\def\tga{\hat{g}^\ast}
\def\trho{\hat{\rho}}
\def\tT{\hat{T}}
\def\tp{\hat{p}}
\def\tu{\hat{u}}
\def\tw{\hat{w}}
\def\tv{\hat{v}}
\def\tva{\hat{v}^\ast}
\def\tbu{\hat{{\bf u}}}
\def\hatdisp{\hat{\eta}}

\def\df1dT{\dfrac{d f_1}{dT}\Bigg|_{\bar{T}}}
\def\df2dT{\dfrac{d f_2}{dT}\Bigg|_{\bar{T}}}

\def\barchi{{\chi}}

\def\d2urhat{\dfrac{d^2 \hat{u}_r}{d r^2}}

\def\kp0{k^{\prime(0)}}

\def\bu{{\bf u}}

\begin{document}
\title[Inviscid stability of compressible flows past compliant surfaces]{Inviscid stability of compressible flows past compliant surfaces}

\author*[1]{\fnm{Mandeep} \sur{Deka}}\email{mandeepdeka@iisc.ac.in}
\author[2]{\fnm{Gaurav} \sur{Tomar}}\email{gtom@iisc.ac.in}
\author[3]{\fnm{Viswanathan} \sur{Kumaran}}\email{kumaran.iisc.ac.in}

\affil*[1,3]{\orgdiv{Department of Chemical Engineering}, \orgname{Indian Institute of Science (IISc) Bangalore}, \orgaddress{\city{Bengaluru}, \postcode{560012}, \state{Karnataka}, \country{India}}}
\affil[2]{\orgdiv{Department of Mechanical Engineering}, \orgname{Indian Institute of Science (IISc) Bangalore}, \orgaddress{\city{Bengaluru}, \postcode{560012}, \state{Karnataka}, \country{India}}}

\abstract{The classical theorems of inviscid stability have been extended for compressible flows past compliant surfaces. We consider normal modes imposed on a plane parallel compressible flow past compliant walls modelled as spring-backed plates and analyze the inviscid equations to derive the theorems. We show that the generalised inflection point criteria of compressible rigid wall flows is modified for flows past dissipative compliant walls. Theorems on the bounds for the wave-speed for unstable modes in the inviscid limit are derived. These are similar to the ones for incompressible compliant wall flows, but are different from compressible rigid wall flows. A new criterion for existence of neutral modes with wave-speeds outside the range of minimum and maximum base velocities is derived for compressible flows past non-dissipative compliant walls. We show that in external compressible flows, neutral modes without a critical point can exist even with dissipative compliant walls, which is not the case in the incompressible limit. }


\maketitle

\section{Introduction}

The study of stability characteristics for flows past compliant surfaces were originally motivated by the prospect of drag reduction in the experiments of \cite{ref-kramer-60,ref-kramer-61,ref-kramer-62}. The seminal works on stability of complaint wall flows were carried out for flat-plate boundary layer by \cite{ref-benjamin-60,ref-benjamin-63}, \cite{ref-landahl-62} and \cite{ref-carpenter-85,ref-carpenter-86}. It was shown that compliant walls produced additional instabilities in the flow, some of which were inviscid in nature as opposed to the Tollmien-Schlichting mode which is a viscous instability. Similar inviscid instabilities were reported for Poiseuille flows as well (see \cite{ref-davies-97,ref-nagata-99,ref-lebbal-2022}). It meant that the Rayleigh inflection point criteria (\cite{ref-rayleigh-1880}), which prohibits the existence of inviscid instabilities in these flows due to the absence of an inflection point, does not hold when the walls are compliant.  

The initial attempts at extending the classical theorems of stability to flows with compliant surfaces were made by \cite{ref-callan-81}. They could extend the Rayleigh and Fj\o{}rtoft's theorems but under very restrictive conditions. The proper extension of some of the classical theorems were done by \cite{ref-yeo-87}. They showed that the critical points of a neutral mode need not coincide with the inflection points unless the compliant walls are dissipative, thereby invalidating the inflection point criteria as an existence condition for inviscid modes in incompressible compliant wall flows. The proper equivalent of the inflection point criteria was later derived by \cite{ref-kumaran-96}, which stated that the required condition for the existence of inviscid instabilities is that the product of the base velocity and its second derivative be negative somewhere in the domain. Although this was derived for flow in a circular channel, it can be easily extended for planar flows (see \cite{ref-kumaran-2021}). From this it becomes clear why a parabolic base flow can become inviscidly unstable in the presence of compliant walls.

The bounds on the wave-speed for unstable modes in the inviscid limit can also be obtained. \cite{ref-yeo-87} derived an equivalent of the semi-circle theorem of \cite{ref-howard-61}, with the lower and upper bounds of the semi-circle denoted by $\baru_L = (\mbox{Min}(\mbox{Min}(\baru),0)$ and $\baru_R = (\mbox{Max}(\mbox{Max}(\baru),0)$, where $\mbox{Min}(\baru)$ and $\mbox{Max}(\baru)$ denote the minimum and maximum of the base velocities, respectively. The same theorem can be extended for incompressible pipe flows (see \cite{ref-kumaran-96}). The Squire's theorem (\cite{ref-squire-33}), that states that two-dimensional modes are always unstable at a lower critical Reynolds number than the three-dimensional modes, cannot be extended for the compliant wall flows, in general. However, \cite{ref-rotenberry-90} extended it for a specific case of channel flow with the wall parameters independent of the perturbation wave-number. It is possible to extend an equivalent of the theorem in the inviscid limit for planar flows as shown in \cite{ref-kumaran-2021} but the same is not possible for cylindrical flows. From a criterion equivalent to the inflection point condition for cylindrical domains, \cite{ref-shankar-2000} showed that a circular pipe flow is stable to axisymmetric perturbations but can be unstable to non-axisymmetric ones.
A comprehensive review, with derivations, of the stability theorems extended for incompressible compliant wall-bounded flows can be found in \cite{ref-kumaran-2021}.

While the stability theorems have been extended to some degree for incompressible compliant wall flows, the same for compressible compliant wall flows have not been attempted. It is important to note that high speed flows generally produce large wall-normal stresses which could lead to deformation even in apparently rigid surfaces. The wave-speed of the inviscid compliant wall modes are generally comparable to the free-surface wave-speeds, which is of the order of $\sqrt{E/\rho}$ ($E$ and $\rho$ being the Young's modulus and density of the solid). For typical metals, this is about $10^3-10^4 m/s$, which can be comparable to the speed of sound in air under certain flow conditions. Therefore, a flow where the Mach number is $O(1)$ past a metal surface can lead to
wall deformation, and the stability of this flow cannot be captured by a rigid wall analysis.
From the perspective of theorems, the extension from incompressible to compressible rigid wall flows is possible for a select few like the semi-circle theorem (see \cite{ref-eckart-63,ref-blumen-1-70,ref-subbiah-90,ref-djordjevic-88,ref-brevdo-93}). Others like the inflection point criterion for the existence of unstable modes can not directly extended for compressible rigid wall flows. Instead, a generalised inflection point criterion, establishing the existence of neutral modes with a critical point, can be derived (see \cite{ref-lees_lin-46,ref-duck-94}). However, this theorem does not hold for neutral modes in an unbounded flow that are supersonic at free-stream (see section \ref{sec:formulation_compliant} for definitions). This is because unlike its incompressible counterpart, compressible flows can allow for neutral modes with non-vanishing eigenfunctions in the free-stream (see appendix \ref{secapp:bc_rayleigh} for details). These non-vanishing eigenfunctions can provide a means for the energy to be carried in or out of the flow, and it is not necessary for the generalised inflection point to coincide with the critical point in the flow. In the presence of dissipative compliant walls, wall dissipation provides another mechanism for energy exchange. Therefore, it is possible to derive certain existence conditions for neutral modes that do not have a counterpart in either rigid wall compressible flows or compliant wall incompressible flows.

In this work, we extend some of the general theorems of stability for compressible flows past compliant walls in the inviscid limit. The governing equation and the wall model is presented in section \ref{sec:formulation_compliant}. In section \ref{sec:inviscidflow_compliant}, the extension of the  stability theorems are derived for general three-dimensional flows. We divide the propositions into three categories; (a) for neutral modes, (b) for non-neutral modes, and (c) for purely span-wise modes. The results and important inferences are summarised in section \ref{sec:summary_compliant}.

\section{Formulation}
\label{sec:formulation_compliant}

The stability of a steady unidirectional compressible shear flow past a compliant surface is studied in the inviscid limit. The base flow is along the $x$ direction, and the temperature and velocity vary in the $y$ direction. General forms for the velocity and temperature profiles, $\baru$ and $\barT$, are used for the base flow. To study the stability, we add small amplitude perturbations ($\mathbf{q}^\prime$) on the base flow ($\bar{\mathbf{q}}$), in the form of normal modes,
\begin{equation}
 \mathbf{q}^\prime = \hat{\mathbf{q}} \exp{\left( \imath \kx (x - c t) + \kz z \right)},
\end{equation} 
where, $\mathbf{q} = [\rho\: u\: v\: w\: T]$ is the general solution vector, with $\hat{\mathbf{q}}$ representing the normal mode amplitude, which is a function of $y$. The stream- and span-wise variation is quantified through the wave-numbers $\kx$ and $\kz$ and the temporal dependence is through the wave-speed $c$. We perform a temporal stability analysis, hence, $\kx, \kz$ are considered to be real while $c$ is complex in general (a positive imaginary part implying an exponentially growing instability in time). The governing equation for the normal modes are derived by subtracting the base flow equations from the compressible Navier-Stokes equations. For the linear analysis, we assume $||\mathbf{q}^\prime||\ll||\bar{\mathbf{q}}||$, hence, all quadratic order terms in the perturbation are ignored in the stability equations. The linearised perturbation equations for a general viscous compressible flow and the corresponding normal mode equations can be found in \cite{ref-deka-2022}. 

In this study, we are concerned with the inviscid limit of the stability equations, for which the normal mode equations reduce to equations \ref{rho_mode_inv}-\ref{p_mode_inv} shown in appendix \ref{secapp:normalmodeeqs_inv}. These equations can be combined to obtain a single evolution equation for the normal velocity mode, known as the compressible Rayleigh equation,
\begin{equation}
 \dfrac{d}{dy}\left(\dfrac{(\baru - c)}{\chi}\dfrac{d\tv}{dy} - \dfrac{1}{\chi}\dfrac{d\baru}{dy}\tv\right) = \dfrac{(\kxsq+\kzsq)(\baru - c)\tv}{\barT} ,
\label{eq:eq35_inv-comp}
\end{equation}
where, 
\begin{equation}
 \chi = \barT - \frac{\Ma^2 \kx^2 (\baru - c)^2}{\kx^2 + \kz^2}. 
 \label{eq:eq36_inv-comp}
\end{equation}
Here, $\Ma$ is the Mach number of the flow. The function $\chi$ can be expressed in terms of the relative Mach number (\cite{ref-mack-84}) as,
\begin{equation}
 \chi = \barT( 1 - \Ma_r^2),
\end{equation} where, 
\begin{equation}
\Ma_r = \frac{\kx\Ma}{\sqrt{\kx^2 + \kz^2}}\left(\frac{\baru - c}{\sqrt{\barT}}\right).
\label{eq:def_ma_r}
\end{equation}
For external flows, the value of relative Mach number in the freestream (far away from sold boundaries where the velocity and temperature are constant) is used to characterize neutrally stable perturbations as subsonic ($\Ma_{r,\infty} < 1$) or supersonic ($\Ma_{r,\infty} > 1$), where the subscript $_\infty$ denotes the free-stream conditions. It can be shown (see \cite{ref-lees_lin-46,ref-mack-84}) that for unbounded flows, supersonic neutral modes do not have vanishing eigenfunctions in the free-stream (see appendix \ref{secapp:bc_rayleigh} for the analysis). This is of importance, since the boundary values of the eigenfunctions are required in deriving the stability theorems. At walls that are rigid, the normal velocity vanishes due to the no penetration condition. However, for compliant walls, the normal velocity at the walls is non-zero, and is related to the normal stress by a suitable wall model.

\begin{figure}
	\begin{center}
 	\includegraphics[scale=0.75]{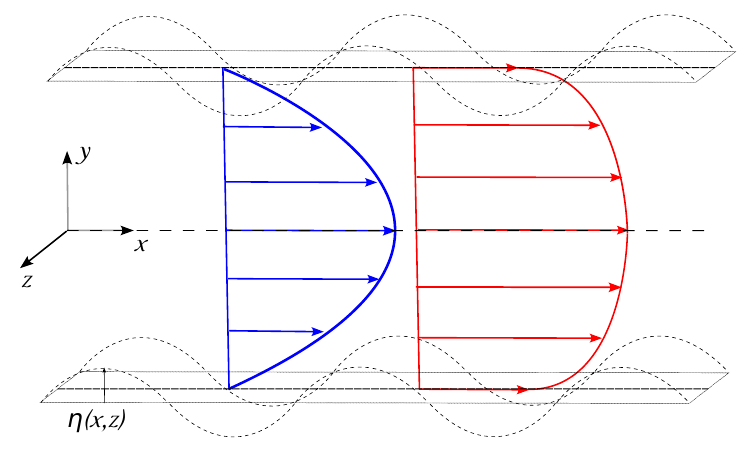}
  	\caption{Schematic of a compressible flow past compliant walls.}
 	\label{fig-comp-1}
	\end{center}
	\end{figure}

The wall model considered in this study is the spring-backed plate model (\cite{ref-benjamin-60,ref-landahl-62,ref-benjamin-63,ref-carpenter-85,ref-carpenter-86,ref-yeo-87,ref-kumaran-2021}). The schematic of a flow past compliant wall is shown in figure \ref{fig-comp-1}. The plate spring is assumed to be perpendicular to the wall in the $y$-direction only. 
The local displacement of the plate from equilibrium along the $y$ direction, denoted by $\disp$, is related to the normal stress exerted by the fluid,
\begin{equation}
\label{compliant_eq}
 I \dfrac{\partial^2 \disp}{\partial t^{2}} + D \dfrac{\partial \disp}{\partial t} + E \disp - T \nabla_s^{2} \disp + B \nabla_s^{4} \disp = \pm \sigma_y ,
\end{equation}
where $\pm \sigma_y$ denotes the component of the total force per unit area in the $y$-direction, exerted by the fluid on the surface. In equation \ref{compliant_eq}, the plus (minus) sign on the right is applicable to a complaint wall with the wall normal into the flow in the direction of positive (negative) $y$. The constants $E$, $T$, $I$, $B$ and $D$ represent non-dimensionalized spring, tension, inertia, bending and damping coefficients, and $\nabla_s$ is the gradient along the surface. For a general unsteady flow, the time derivative of $\disp$ is the normal velocity at the surface, therefore, equation \ref{compliant_eq} can be used to relate the normal velocity to the normal stress at the surface.

For a steady parallel base flow, the flow variables do not vary along the stream- or span-wise directions, implying that the normal stress is constant across the compliant wall. Therefore, at the base flow the normal stresses at the compliant surface can produce a spatially constant displacement of the surface in the wall normal direction. Hence, for the purpose of the stability analysis, the base flow can be considered with the initial pre-stressed state of the compliant surface. With the normal modes imposed on the base flow, the normal displacement produced can be represented as a normal mode,
\begin{equation}
\label{disp_pert_eq}
 \disp = \hatdisp \exp\left({\imath \kx (x - ct) + \imath \kz z}\right).
\end{equation}
Substituting the above displacement into equation \ref{compliant_eq}, the evolution for $\hatdisp$ is,
\begin{equation}
 \left( -\kxsq c^2 I - \imath \kx c D + E + \left( \kxsq + \kzsq \right)T + \left( \kxsq + \kzsq \right)^2 B \right) \hatdisp  = \pm \hat{\sigma}_y.
\label{eq:comp_wall_spd}
\end{equation}
where, $\hat{\sigma}_y$ denotes the normal mode corresponding to the $y$-direction force per unit area at the wall. In the absence of viscous stresses, the normal force per unit area acted upon by the fluid on the surface is the fluid pressure, therefore,
\begin{equation}
 \hat{\sigma}_y = \frac{\tp}{\gamma \Ma^2}\Big|_{y = y_w},
\label{eq:normal_stress_rel_inv}
\end{equation} 
where $y_w$ is the location of the wall. The time derivative of the normal displacement is equal to the normal velocity at the wall, therefore, 
\begin{equation}
\label{disp_vel_rel}
 \hat{v}|_{y=y_w} = -\imath \kx c \:\hatdisp.
\end{equation}
Substituting equations \ref{eq:normal_stress_rel_inv} and \ref{disp_vel_rel} in equation \ref{eq:comp_wall_spd}, a relation between pressure and the normal velocity modes can be obtained at the surface,
\begin{equation}
\label{eq:compl_wl_disp_mode_rel}
 \pm \frac{\tp}{\gamma \Ma^2}\Big|_{y = y_w}  = \dfrac{\imath ( E + \left( \kxsq + \kzsq \right)T + \left( \kxsq + \kzsq \right)^2 B - \imath \kx c D - \kxsq c^2 I )}{\kx c}  \hat{v}|_{y=y_w}  .
\end{equation}

\section{Results}
\label{sec:inviscidflow_compliant}

In this section, we derive some general results for stability of an inviscid compressible flow past
a compliant surface. In general, the following types of parallel flow configurations can be considered,
\begin{enumerate}
\item An internal flow bounded by at least one compliant wall (example, a compressible Couette or Poiseuille flow).
\item An external flow past a compliant wall (example, boundary layer flows). 
\end{enumerate}
At a compliant surface, the normal velocity is related to the pressure in equation \ref{eq:compl_wl_disp_mode_rel}. 
However, to derive the theorems, it is convenient to replace the pressure by the normal velocity and its derivatives at the compliant surface. This is carried out by deriving an evolution equation for the pressure, by adding $\barT \times$ equation \ref{rho_mode_inv} (appendix \ref{secapp:normalmodeeqs_inv}) to equation \ref{T_mode_inv} (appendix \ref{secapp:normalmodeeqs_inv}). Using equation \ref{p_mode_inv}, the resulting equation is,
\begin{equation}
 \imath \kx (\baru - c) \tp + \gamma \left(\dfrac{d\tv}{dy} + \imath \kx \tu + \imath \kz \tw \right) = 0 .
\label{eq:eq41_inv-comp}
\end{equation}
Adding equation \ref{u_mode_inv} to ${\kz}/{\kx} \times$ equation \ref{w_mode_inv}, we can express $(\kx\tu + \kz\tw)$ in terms of $\tv$ and $\tp$. Substituting into equation \ref{eq:eq41_inv-comp}, we get the relation,
\begin{equation}
 \dfrac{d\tv}{dy} = \dfrac{1}{(\baru - c)}\dfrac{d\baru}{dy}\tv + \dfrac{\imath \chi (\kxsq + \kzsq)}{\gamma \Ma^2 \kx (\baru - c)}\tp.
\label{eq:eq43_inv-comp}
\end{equation} 
Substituting $\tp$ from equation \ref{eq:compl_wl_disp_mode_rel} into the right hand side of the above equation, we get the relation at the compliant surface,
\begin{equation}
\begin{aligned}
\dfrac{d\tv}{dy}&\Big|_{y=y_w} = \Bigg( -\dfrac{1}{c}\dfrac{d\baru}{dy}\Big|_{y=y_w} \pm \dfrac{\chi|_{y=y_w} (\kxsq + \kzsq) ( E_{eq} - \imath \kx c D - \kxsq c^2 I )}{ \kxsq c^2} \Bigg) \tv|_{y=y_w},
\end{aligned}
\label{eq:eq43a_inv-comp}
\end{equation}
where, $E_{eq} = E + \left( \kxsq + \kzsq \right)T + \left( \kxsq + \kzsq \right)^2 B$. The above relation assumes $\baru|_{y=y_w} = 0$, i.e, the steady parallel flow considered is corresponding to a stationary compliant wall in the base state. Note that the analysis shown in this study does not extend for compliant walls with tangential motion at the base state, for example a Couette flow with the moving wall being compliant instead of the rigid wall. However, this is the case even in the incompressible limit, where all theorems derived are for stationary compliant walls only. 

\subsection{Theorems for neutral ($\cI = 0$) modes.}
\label{sec:theorems}

The extension of the Rayleigh inflection point theorem for compressible flows past compliant walls can be derived. We first multiply equation \ref{eq:eq35_inv-comp} by $\tva$, the complex conjugate of $\tv$, divide by $\baru - c$, and simplify, to obtain the relation,
 \begin{eqnarray}
 \frac{d}{d y} \left( \frac{\tva}{\chi} \frac{d \tv}{d y} \right) = \dfrac{1}{\chi}\left| \dfrac{d\tv}{dy} \right|^2 + \frac{\left|\tv\right|^2}{(\baru - c)} \frac{d}{d y} \left( \frac{1}{\chi} \frac{d \baru}{d y} \right) + \frac{(\kxsq + \kzsq) \left|\tv\right|^2}{\barT}. 
\label{eq:eq42aa_inv-comp}
 \end{eqnarray}
 This equation is multiplied by the wave-speed $c$ and integrated across the domain of flow $y \in [y_l,y_h]$, to obtain,
 \begin{equation}
  \left( \frac{c \tva}{\chi} \frac{d \tv}{d y} \right)\Bigg|_{y_l}^{y_h} = c \int_{y_l}^{y_h} dy \left(\dfrac{1}{\chi}\left| \dfrac{d\tv}{dy} \right|^2 + \frac{\left|\tv\right|^2}{(\baru - c)} \frac{d}{d y} \left( \frac{1}{\chi} \frac{d \baru}{d y} \right) + \frac{(\kxsq + \kzsq) \left|\tv\right|^2}{\barT} \right) .
\label{eq:eq42a_inv-comp}
 \end{equation}
For a compliant surface, ${d \tv}/{d y}$ from equation \ref{eq:eq43a_inv-comp} is substituted into the left hand side of the above equation. The imaginary part of the resulting equation, after simplification, can be written as,
\begin{equation}
\begin{aligned}
 &\left(\dfrac{2 \kx^2 \Ma^2 \cR \cI}{(\kxsq + \kzsq)|\chi|^2}\dfrac{d\baru}{dy}|\tv|^2\right)\Bigg|_{y_l}^{y_h} - \sum_{y=y_l,y_h} \left( \left(\dfrac{\kxsq+ \kzsq}{\kxsq}\right) \left( \left(\dfrac{E_{eq}}{|c|^2} + \kx I \right) \cI + \kx D \right) |\tv|^2\right)\\
 &\:\:= \int_{y_l}^{y_h} dy \:\Bigg\{ \dfrac{\cI \left(\barT - \frac{\kxsq \Ma^2 }{\kxsq + \kzsq}\left((\baru - \cR)^2 - \cI^2 + 2(\baru - \cR)\cR\right)\right)}{|\chi|^2}\left| \dfrac{d\tv}{dy} \right|^2 \\
&\:\:\:\:\:\:\:\:\:\:\:\:\:\:+ \dfrac{\cI \baru |\tv|^2}{|\baru - c|^2}\dfrac{d}{dy}\left( \dfrac{\barT - \frac{\kxsq \Ma^2 }{\kxsq + \kzsq} \left((\baru - \cR)^2 - \cI^2 \right)}{|\chi|^2}\dfrac{d\baru}{dy} \right) \\
&\:\:\:\:\:\:\:\:\:\:\:\:\:\:\:\:\:\:\:- \dfrac{((\baru - \cR)\cR + \cI^2) |\tv|^2}{|\baru - c|^2}\dfrac{d}{dy}\left( \dfrac{2\kxsq\Ma^2(\baru - \cR) \cI}{(\kxsq+\kzsq)|\chi|^2}\dfrac{d\baru}{dy} \right) + \dfrac{\cI(\kxsq+\kzsq)|\tv|^2}{\barT} \Bigg\}.
\end{aligned}
\label{eq:eq42b_inv-comp}
\end{equation}
Note that in the left hand side of the above equation, both boundaries are assumed to be compliant (summation is implied across both boundaries). If only one wall is compliant, then the terms on the left hand side are evaluated only at the compliant boundary. In that case, the contribution to the boundary term from the non-compliant boundary is identically zero, provided the non-compliant boundary is either a rigid wall (as in case of an internal flow with one compliant wall) or is located at the freestream (for external flows), where the mode in consideration is a neutral subsonic mode (see appendix \ref{secapp:bc_rayleigh}).

In equation \ref{eq:eq42b_inv-comp}, the only term that does not decrease to
zero for $c_I \rightarrow 0$ is the boundary term proportional to \( D \) on the left side. All the integral terms on the right hand side are regular if $\cR$ is outside the range of $(\mbox{Min}(\baru),\mbox{Max}(\baru))$. If we restrict attention to non-dissipative walls ($D=0$), for modes with $\cI \rightarrow 0$, the boundary terms on left hand side of equation \ref{eq:eq42b_inv-comp} are $ O(|c_I|)$. The terms on the right are also $O(c_I)$ if $\cR$ is outside the range
$(\mbox{Min}(\baru),\mbox{Max}(\baru))$. If $\cR$ is in the range $(\mbox{Min}(\baru),\mbox{Max}(\baru))$, the some of
the terms on the right in equation \ref{eq:eq42b_inv-comp} are singular for $|\baru - c_R| = 0$, and the terms
have to be evaluated for $c_I \ll 1$ and $|\baru - c_R| \sim O(|c_I|)$. The first and fourth terms in the flower bracket
on the right are $O(c_I)$. The third term on the right is $O(1)$ in this limit, and therefore provides a vanishing
contribution to the integral. The largest contribution to the integral could only be due to the second
term on the right, which is proportional to $(c_I/|\baru - c_R|^2)$ in the limit $c_I \rightarrow 0$. Since all
the other terms in the integrand on the right are $O(1)$ or smaller, it is necessary that the second term
on the right also be $O(1)$ or smaller for $|\baru - c_R| \sim c_I \rightarrow 0$. This implies that,
\begin{equation*} \frac{d}{dy} \left( \frac{1}{\barT} \frac{d \baru}{d y} \right) \sim O(|c_I|). \end{equation*}
This leads to the following result,
\begin{theoremcompliant}
A neutral mode for an internal flow or a subsonic neutral mode for an external flow, with $\mbox{Min}(\baru) < c_R < \mbox{Max}(\baru)$, can exist in the inviscid limit, in a compressible flow past non-dissipative compliant surfaces ($D=0$) only if,
 \[
  \frac{d}{dy} \left( \frac{1}{\barT} \frac{d \baru}{d y} \right) = 0 ,
 \]
at the location where $\baru = c_R$.
\label{theorem:proposition1_compliant}
\end{theoremcompliant}
Here, we have substituted $\chi = \bar{T}$ based on equation \ref{eq:eq36_inv-comp} in the limit $|\baru - c_R| \ll 1$.
This result is the generalised inflection point theorem (GIP) for compressible flows past rigid surfaces (see \cite{ref-lees_lin-46,ref-duck-94}). Therefore, the existence condition for neutral modes with a critical point in the flow 
past a non-dissipative surface is the same as that for an incompressible flow past non-dissipative compliant walls (see \cite{ref-kumaran-2021}).

For neutral modes in internal flows or subsonic neutral modes in external flows, it is evident from equation \ref{eq:eq42b_inv-comp} that the left hand side is non-zero {and negative} when $D \neq 0$. This can only be balanced by a contribution from the singular part of the integrand which is the second term in the flower brackets on the right hand side. This term has a critical point $y_c$, where $\baru(y_c) = \cR$. In the limit of $\cI \rightarrow 0$, we can simplify the integral in terms of the jump across the critical point,
\begin{equation*}
 \int_{y_l}^{y_h} dy \left( \cdot \right) = \int_{y_c^- }^{y_c^+} dy \left( \cdot \right)_{y=y_c} + O(\cI).\:\:
\end{equation*} 
{Here, it is assumed that the critical point is in the interior of the domain at not located at $y_l$ or $y_h$.}
The dominant term under the integral on the right hand side is the second term, which can be simplified as,
\begin{equation}
\begin{aligned}
 \lim_{\cI \rightarrow 0} \int_{y_c^-}^{y_c^+} dy &\left(\left(|\tv|^2 \baru \dfrac{d}{dy}\left( \dfrac{1}{\barT}\dfrac{d\baru}{dy} \right)\right)\Bigg|_{y=y_c}\dfrac{\cI}{(\baru - \cR)^2 + \cI^2}\right) \\
&= {\left(|\tv|^2 \frac{\baru}{\baru^\prime} \dfrac{d}{dy}\left( \dfrac{1}{\barT}\dfrac{d\baru}{dy} \right)\right)\Bigg|_{y=y_c}} \lim_{\cI \rightarrow 0} \int_{\baru(y_c^-)}^{\baru(y_c^+)} \dfrac{\cI d\baru}{(\baru - \cR)^2 + \cI^2} ,\\
\end{aligned}
\label{eq:jump_crit_pt_compliant}
\end{equation}
where, $\baru^\prime = (d \baru/d y)$, which is assumed to be finite at the critical point, and we have used the substitution, $dy = d\baru/\baru^\prime_c $. The limit on the right hand side can be evaluated,
\begin{equation}
\begin{aligned}
 \lim_{\cI \rightarrow 0} \int_{\baru(y_c^-)}^{\baru(y_c^+)} \dfrac{\cI d\baru}{(\baru - \cR)^2 + \cI^2}\: &= \lim_{\cI \rightarrow 0} \left(\tan^{-1}\left(\frac{\baru - \cR}{\cI}\right)\right)\Bigg|_{\baru(\yc^-)}^{\baru(\yc^+)} \\&=
											\:\begin{cases} 
											\:\:\:\pi, &\text{if $\baru^\prime|_{y=y_c}>0$} \\
											-\pi, &\text{if $\baru^\prime|_{y=y_c}<0$}
											\end{cases} \:\:\equiv\:\: \pi \left(\frac{\baru^\prime}{|\baru^\prime|}\right)\Bigg|_{y=y_c}.
\end{aligned}
\end{equation}
Substituting the above relation into equation \ref{eq:jump_crit_pt_compliant} and \ref{eq:eq42b_inv-comp} in the limit of $\cI \rightarrow 0$ and $\mbox{Min}(\baru) < c_R < \mbox{Max}(\baru)$, the equation \ref{eq:eq42b_inv-comp} (upto an $O(\cI)$) simplifies as,
\begin{equation}
  - \sum_{y=y_l,y_h} \left(\dfrac{(\kxsq+\kzsq) D |\tv|^2}{\kx}\right) = \pi\left(\frac{|\tv|^2}{|\baru^\prime|}\right)\Bigg|_{y = y_c} \left( \baru \dfrac{d}{dy}\left( \dfrac{1}{\barT}\dfrac{d\baru}{dy} \right)\right)\Bigg|_{y=y_c}.
\end{equation}
Since $D>0$ always, the right hand side must be negative. This leads to the following result,
\begin{theoremcompliant}
A neutral mode for an internal flow or a subsonic neutral mode for an external flow, with $\mbox{Min}(\baru) < c_R < \mbox{Max}(\baru)$, can exist in the inviscid limit, in a compressible flow past a dissipative compliant surface ($D > 0$) only if,
  \begin{equation*}
  \baru\:\dfrac{d}{dy}\left( \dfrac{1}{\barT}\dfrac{d\baru}{dy} \right) < 0 , 
 \end{equation*}
 at the location where $\baru = c_R$.
\label{theorem:proposition1b_compliant}
\end{theoremcompliant}
The above theorem is valid for an internal flows with a one or both dissipative walls. For external flows with supersonic neutral modes, both propositions \ref{theorem:proposition1_compliant} and \ref{theorem:proposition1b_compliant} do not hold as the boundary term in equation \ref{eq:eq42a_inv-comp} has a non-zero imaginary part (see appendix \ref{secapp:bc_rayleigh}). For such modes, it is understood that the energy generated (or consumed) due to the phase shift at the critical point is brought in (or carried out) by the neutral waves at freestream. In the presence of a dissipative compliant wall, some of the energy can be dissipated through the wall, however, the relative magnitudes are dependent on the solution and hence, a universal criterion cannot be determined.

Propositions \ref{theorem:proposition1_compliant} and \ref{theorem:proposition1b_compliant} hold only for neutral modes with $\cR$ in the range of $(\minbaru,\maxbaru)$. But neutral modes with $\cR$ outside this range exist in all compressible shear flows (generally referred to as higher modes) and are known to determine the stability of these flows at finite Mach numbers (see \cite{ref-mack-87,ref-duck-94,ref-deka-2023}). We now derive a condition for the existence of these type of modes.
Defining the function $\tg$, 
\begin{equation}
\tg = \frac{\tv}{(\baru - c)},
\end{equation}
and rewriting the Rayleigh equation, equation \ref{eq:eq35_inv-comp}, is terms of $\tg$, we get,
\begin{eqnarray}
 \frac{d}{d y} \left( \frac{(\baru - c)^2}{\chi} \frac{d \tg}{d y} \right) & = & \frac{(\kxsq + \kzsq) (\baru - c)^2 \tg}{\barT} .
\end{eqnarray}
This equation is multiplied by the complex conjugate $\tga$ and integrated across the channel,
\begin{eqnarray}
 \left. \left( \frac{(\baru - c)^2 \tga}{\chi} \frac{d \tg}{d y} \right) \right|_{y_l}^{y_h} -
 \int_{y_l}^{y_h} dy \left( \frac{(\baru - c)^2}{\chi} \left| \frac{d \tg}{d y} \right|^2
 + \frac{(\kxsq + \kzsq) (\baru - c)^2 |\tg|^2}{\barT} \right) & = & 0. 
 \label{eq:eq37_inv-comp}
\end{eqnarray}
To obtain the derivative of $\tg$ at the compliant wall, we substitute $\tv = (\baru - c) \tg$ into equation \ref{eq:eq43_inv-comp} and obtain a relation between $\tg$ and $\tp$. Then we substitute $\tp$ from equation \ref{eq:compl_wl_disp_mode_rel} at the compliant wall, which leads to the boundary condition,
\begin{equation}
 \dfrac{d\tg}{dy}\Big|_{y=y_w} = \pm \dfrac{\chi|_{y=y_w} (\kxsq + \kzsq) ( E_{eq} - \imath \kx c D - \kxsq c^2 I )}{\kxsq c^2 }\tg|_{y=y_w} ,
\label{eq:eq45_inv-comp}
\end{equation}
where, $E_{eq} = E + \left( \kxsq + \kzsq \right)T + \left( \kxsq + \kzsq \right)^2 B$. Substituting the above into the left hand side of equation \ref{eq:eq37_inv-comp}, we obtain,
\begin{equation}
\begin{aligned}
\sum_{y=y_l,y_h} &\dfrac{(\kxsq + \kzsq) ( E_{eq} - \imath \kx c D - \kxsq c^2 I )}{ \kxsq }|\tg|^2  \\
	&\hspace{1.5cm}= \int_{y_l}^{y_h} dy \left( \frac{(\baru - c)^2}{\chi} \left| \frac{d \tg}{d y} \right|^2
 + \frac{(\kxsq + \kzsq) (\baru - c)^2 |\tg|^2}{\barT} \right) .
\end{aligned}
\label{eq:eq64_inv-comp}
\end{equation}
The summation on the left side is over the compliant walls. If one boundary is rigid or at infinity for a subsonic neutral mode, the contribution can be shown to be identically zero due to the zero normal velocity condition. Therefore, in that case, considering the imaginary part of equation \ref{eq:eq64_inv-comp}, we get $D = 0$. However, for an external flow with a supersonic neutral mode, the left hand side of equation \ref{eq:eq37_inv-comp} has a non-zero imaginary part that depends on the relative energies of the incoming and outgoing waves at infinity that can balance against the wall dissipation term (see equation \ref{eqapp:eq63} in appendix \ref{secapp:bc_rayleigh}). Thus, we get the following result, 
\begin{theoremcompliant}
 A neutral mode with $\cR$ outside the range of $(\minbaru,\maxbaru)$ can only exist in the inviscid limit, in an external flow with a dissipative compliant wall ($D > 0$), if the mode is supersonic at freestream. A neutral mode with $\cR$ outside the range of $(\minbaru,\maxbaru)$ cannot exist in an internal flow with a dissipative wall in the inviscid limit.
\label{theorem:proposition5b_compliant}
\end{theoremcompliant}

For non-dissipative wall ($D = 0$), defining the wave-speed ($\co$) of the unforced harmonic oscillations of the wall,
\begin{equation}
 \co = \dfrac{1}{\kx}\sqrt{\dfrac{E_{eq}}{I}} ,
\end{equation}
equation \ref{eq:eq64_inv-comp} can be re-written as,
\begin{equation}
\sum_{y=y_l,y_h} (\kxsq + \kzsq) I ( \co^2 - c^2 )|\tg|^2 = \int_{y_l}^{y_h} dy \left( \frac{(\baru - c)^2}{\chi} \left| \frac{d \tg}{d y} \right|^2 + \frac{(\kxsq + \kzsq) (\baru - c)^2 |\tg|^2}{\barT} \right) .
\label{eq:eq66_inv-comp}
\end{equation}
It is clear that for neutral modes ($\cI = 0$), the left hand side is negative if $c^2 > \co^2$, and all the terms on the right side 
inside the integral to be positive with the exception of $\chi$. This leads to the following result,
 \begin{theoremcompliant}
 For a flow past non-dissipative ($D=0$) compliant walls, an neutral mode in an internal flow or a subsonic neutral mode in an external flow with $\cR$ outside the range of $(\mbox{Min}(\baru),\mbox{Max}(\baru))$ and $\kx\cR > \sqrt{{E_{eq}}/{I}}$ can exist only if 
 \begin{eqnarray}
  \chi = \barT - \frac{\Ma^2 \kx^2 (\baru - c)^2}{\kx^2 + \kz^2}
 \end{eqnarray}
 is negative somewhere in the domain.
 \label{theorem:proposition5_compliant}
 \end{theoremcompliant}
There is no theorem equivalent to proposition \ref{theorem:proposition5b_compliant} for either incompressible flows past compliant walls or compressible flows past rigid walls. However, proposition \ref{theorem:proposition5_compliant} is equivalent to the $\chi$-criteria (see \cite{ref-deka-2022,ref-deka-2023}) obtained for internal compressible flows past rigid walls but with an added constraint, $\kx\cR>\sqrt{{E_{eq}}/{I}}$. This theorem also has no counterpart for incompressible flows past compliant walls.

\subsection{Theorems of stable/unstable ($\cI \neq 0$) modes.}

For non-neutral modes, the velocity perturbations have to vanish at rigid walls for internal flows as well as at infinity for external flows, for all modes. Therefore, the results derived here will hold for both internal and external flows with at least one compliant wall. To derive bounds on the wave-speed for non-neutral modes, we rewrite the integral,
\begin{eqnarray}
\int_{y_l}^{y_h} dy \left( \frac{(\baru - c)^2}{\chi} \left| \frac{d \tg}{d y} \right|^2
 + \frac{(\kxsq + \kzsq) (\baru - c)^2 |\tg|^2}{\barT} \right) 
 = \int_{y_l}^{y_h} dy \left( \Psi - (\baru - c)^2 \Phi  \right) 
 \label{eq:eq38_inv-comp}
\end{eqnarray}
where the functions
\begin{eqnarray}
 \Psi & = & \frac{\Masq \kxsq |\baru - c|^4}{(\kxsq + \kzsq) |\chi|^2} \left| \frac{d \tg}{d y} \right|^2, \: \:
 \Phi = \left(\frac{\barT}{|\chi|^2} \left| \frac{d \tg}{d y} \right|^2 + \frac{(\kxsq + \kzsq) |\tg|^2}{\barT} \right),
 \label{eq:eq38a_inv-comp}
\end{eqnarray}
are positive throughout the domain. Substituting the derivative of $\tg$ at the boundary from equation \ref{eq:eq45_inv-comp}, and using equation \ref{eq:eq38_inv-comp}, we can write equation \ref{eq:eq37_inv-comp} as,
\begin{equation}
\begin{aligned}
\sum_{y=y_l,y_h} &\dfrac{(\kxsq + \kzsq) ( E_{eq} - \imath \kx c D - \kxsq c^2 I )}{\kxsq }|\tg|^2 +  \int_{y_l}^{y_h} dy \left( \Psi - (\baru - c)^2 \Phi  \right) = 0 .
\end{aligned}
\label{eq:eq46_inv-comp}
\end{equation}
If we take the imaginary part of equation \ref{eq:eq46_inv-comp}, we obtain,
\begin{equation}
 2 \cI \int_{y_l}^{y_h} dy (\baru - \cR) \Phi = \sum_{y_l,y_h} \dfrac{\cR (\kxsq + \kzsq)(D + 2 \cI \kx I)|\tg|^2}{\kx}.
\label{eq:eq47_inv-comp}
\end{equation}
For unstable modes with $\cI>0$, the right side of the above equation has the
same sign as $\cR$. For downstream traveling waves with $\cR > 0$, the right side
is positive. For the left side to be positive, it is necessary that the maximum
of $\baru$ is positive, and $\cR$ is less than the maximum of $\baru$. Similarly, for $\cR < 0$,
the right side is negative. The left side is negative only if the minimum of
$\baru$ is less than 0, and $\cR$ is greater than the minimum of $\baru$. From this, we
obtain the following result,
\begin{theoremcompliant}
In the inviscid limit, unstable modes ($\cI > 0$) can exist in a compressible shear flow past compliant walls if, 
\begin{equation}
\mbox{Min}(\mbox{Min}(\baru), 0) < \cR < \mbox{Max}(\mbox{Max}(\baru), 0).
\end{equation}
\label{theorem:proposition2_compliant}
\end{theoremcompliant}
This theorem is same as the one obtained in the incompressible limit (\cite{ref-yeo-87,ref-kumaran-2021}).

For non-dissipative compliant walls ($D = 0$), it is clear from equation \ref{eq:eq47_inv-comp} that modes with $\cR$ outside the range of $(\mbox{Min}(\mbox{Min}(\baru), 0),\mbox{Max}(\mbox{Max}(\baru), 0))$ can only exist if $\cI = 0$. For $D \neq 0$, we can divide equation \ref{eq:eq47_inv-comp} by $\cI$ and rearrange the terms to obtain the relation,
\begin{equation}
 \left( \int_{y_l}^{y_h} dy \: (\baru - \cR) \Phi - \sum_{y_l,y_h} {\cR (\kxsq + \kzsq) I |\tg|^2}\right) = \sum_{y_l,y_h} \dfrac{\cR (\kxsq + \kzsq)D |\tg|^2}{2\cI\kx}.
\label{eq:eq47_inv-comp-2}
\end{equation}
For $\cR > \mbox{Max}(\mbox{Max}(\baru), 0)$, the left hand side of the above equation is always less than zero, therefore from the right hand side, so is $\cI$. Similarly, for $\cR < \mbox{Min}(\mbox{Min}(\baru), 0)$, the left hand side is always greater than zero, but since $\cR$ is less than zero, from the right hand side, $\cI$ must again be less than zero. This gives the following result,
\begin{theoremcompliant}
For compressible shear flows past compliant walls, modes with $\cR$ outside the range of $(\mbox{Min}(\mbox{Min}(\baru), 0),\mbox{Max}(\mbox{Max}(\baru), 0))$ are always stable ($\cI < 0$) in the inviscid limit, if at least one wall is dissipative ($D>0$), otherwise they are neutrally stable ($\cI = 0$).
\label{theorem:proposition2b_compliant}
\end{theoremcompliant}
 
If we multiply equation \ref{eq:eq46_inv-comp} by $\ca$ and take the imaginary part, we obtain,
\begin{equation}
\begin{aligned}
 \sum_{y=y_l,y_h} &\dfrac{\left({\kxsq + \kzsq}\right)\left( \cI \left( E_{eq} + \kxsq |c|^2 I\right) + \kx |c|^2 D\right)|\tg|^2}{{\kxsq}} + \cI \int_{y_l}^{y_h} dy \left( \Psi + (|c|^2 - \baru^2)\Phi\right) = 0 .
\end{aligned}
\label{eq:eq49_inv-comp}
\end{equation}
For unstable modes with $\cI > 0$, the first term on the left is always positive.
The second term can be negative only if $|c|^2 < \baru^2$ somewhere in the flow. For non-dissipative walls ($D=0$) the first terms has the same sign as $\cI$ while the second term could has the opposite sign if only $|c|^2 < \baru^2$ somewhere in the flow. This leads to the following result,

\begin{theoremcompliant}
In the inviscid limit, unstable modes ($\cI>0$) can exist for dissipative walls ($D > 0$), or non-neutral modes ($\cI \neq 0$) can exist for non-dissipative walls ($D = 0$) only if, 
\begin{equation}
 |c|^2 < \mbox{Max}(\baru^2) .
\end{equation}
\label{theorem:proposition3_compliant}
\end{theoremcompliant}

For non-neutral modes ($\cI \neq 0$), equations \ref{eq:eq47_inv-comp} and \ref{eq:eq49_inv-comp} can be rewritten as,
\begin{equation}
\begin{aligned}
  \int_{y_l}^{y_h} dy \baru \Phi = & \:	c_R \int_{y_l}^{y_h} dy \Phi + \sum_{y_l,y_h} \dfrac{\cR (\kxsq + \kzsq)(D + 2 \cI \kx I)|\tg|^2}{2 \cI \kx},\\
  \int_{y_l}^{y_h} dy \baru^2 \Phi = & \int_{y_l}^{y_h} dy ((c_R^2 + c_I^2) \Phi + \Psi) + \sum_{y_l,y_h} \dfrac{(\kxsq + \kzsq)\left( \cI \left( E_{eq} + \kxsq |c|^2\right) + \kx |c|^2 D\right)|\tg|^2}{\kxsq }.
\end{aligned}
\label{eq:eq51_inv-comp}
\end{equation}
The base velocity field satisfies the identity
\begin{eqnarray}
\int dy (\baru - \mbox{Min}(\baru))(\baru - \mbox{Max}(\baru)) \Phi & < & 0, \label{eq:eq52_inv-comp}
\end{eqnarray}
where $\mbox{Min}(\baru)$ and $\mbox{Max}(\baru)$ are the minimum and maximum values of the mean velocity.
Substituting equation \ref{eq:eq51_inv-comp} for the integrals of the $\baru \Phi$ and
$\baru^2 \Phi$, we obtain,
\begin{equation}
\begin{aligned}
 \int_{y_l}^{y_h} dy &[((c_R^2 + c_I^2) - c_R (\mbox{Min}(\baru) + \mbox{Max}(\baru)) + \mbox{Min}(\baru) \mbox{Max}(\baru)) \Phi + \Psi] \\
	&+ \sum_{y_l,y_h} \dfrac{(\kxsq + \kzsq)}{\kxsq}\Big( E_{eq} + \kxsq \left( \cR^2 + \cI^2 - \cR (\mbox{Min}(\baru) + \mbox{Max}(\baru)) \right) \Big)\\
	&\:\:\:\:\:\:\:\:\:\:\:\:+ \sum_{y_l,y_h} \dfrac{(\kxsq + \kzsq) D \left( 2(c_R^2 + c_I^2) - c_R (\mbox{Min}(\baru) + \mbox{Max}(\baru))\right)}{2 \cI \kx} < 0. 
\end{aligned}
\label{eq:eq53_inv-comp}
\end{equation}
Using this, it can be shown, by separately considering $0 < \mbox{Min}(\baru) < \mbox{Max}(\baru)$ ,
$\mbox{Min}(\baru) < 0 < \mbox{Max}(\baru)$ and $\mbox{Min}(\baru) < \mbox{Max}(\baru) < 0$, that
 \begin{theoremcompliant}
 For a non-neutral mode (\( c_I \neq 0 \)) in a compressible shear flow past compliant walls, in the inviscid limit,
 \begin{eqnarray}
  (c_R - \mbox{$\frac{1}{2}$} (\baru_U + \baru_L))^2 + c_I^2 & < & \mbox{$\frac{1}{4}$} (\baru_U - \baru_L)^2,
  \label{eq:eq54_inv-comp}
 \end{eqnarray}
 where $\baru_U = \mbox{Max}(\mbox{Max}(\baru),0)$ and $\baru_L = \mbox{Min}(\mbox{Min}(\baru),0)$.
 \label{theorem:proposition4_compliant}
 \end{theoremcompliant} 
This is equivalent to the semi-circle theorem which has the same form for incompressible flows past compliant walls (\cite{ref-yeo-87,ref-kumaran-2021}).

\subsection{Theorems for purely span-wise perturbations ($\kx = 0$).}

For span-wise perturbations imposed on the mean flow, the density, velocity and temperature
are expressed as,
 \begin{eqnarray}
 \rho & = & \barrho + \trho \exp{(\imath \kz z + s t))}, \label{eq:span1_comp} \\
 \bu & = & \baru {\bf e}_x + \tbu \exp{(\imath \kz z + s t)}, \label{eq:span2_comp} \\
 T & = & \barT + \tT \exp{(\imath \kz z + s t)}, \label{eq:span3_comp} \\
 p & = & \barp + \tp \exp{(\imath \kz z + s t)}. \label{eq:span4_comp}
\end{eqnarray}
Here, $s$ is complex in general,and the real part of $s$ is the exponential growth rate in time. The above equations can be simplified to express the density and velocity in terms of the temperature,
\begin{eqnarray}
 \tu & = & \frac{1}{\barrho \gamma \Ma^2 s^2} \frac{d \baru}{d y} \frac{d \tp}{d y},\: \:  \tv = - \frac{1}{\barrho \gamma \Ma^2 s} \frac{d \tp}{d y},\: \: \tw = - \frac{\imath \kz \tp}{\barrho \gamma
 \Ma^2 s}, \label{eq:span5_comp} \\
 \tT & = & \frac{(\gamma-1) \tp}{\barrho \gamma} + \frac{1}{\barrho \gamma \Ma^2 s^2} \frac{d \barT}{d y} 
 \frac{d \tp}{d y}, \: \: \trho = \frac{\tp}{\gamma \barT} - \frac{1}{\gamma \barT \Ma^2 s^2} \frac{d \barT}{d y}
 \frac{d \tp}{d y}. \label{eq:span6_comp}
\end{eqnarray}
These are inserted into the mass conservation equation \ref{rho_mode_inv} to obtain one second order equation for 
the pressure,
\begin{eqnarray}
 \frac{d}{d y} \left( \barT \frac{d \tp}{d y} \right) - (\Ma^2 s^2 + \barT \kz^2) \tp & = & 0.
 \label{eq:span7_comp}
\end{eqnarray}
For the compliant wall, the wall displacement in the normal direction can be written in the form,
\begin{equation}
 \disp = \hatdisp \exp{\left( \imath \kx z + s t \right)}.
 \label{eq:span8_comp}
\end{equation}
Since rate of change of the normal displacement is equal to the normal velocity at the walls, the
normal velocity at the wall can be expressed as $\tv|_{y=y_w} = s \hatdisp$.
Substituting these relations into the wall equation \ref{compliant_eq}, we obtain the relation in the inviscid limit,
\begin{equation}
 \frac{\tp}{\gamma\Ma^2}\Big	|_{y=y_w} = \pm \dfrac{\left(E + T \kzsq + B \kz^4 + D s + I s^2\right) }{s} \tv|_{y=y_w} .
 \label{eq:span10_comp}
\end{equation}
Multiplying equation \ref{eq:span7_comp} with $\tp^*$ and integrating across the width of the channel, we obtain the expression,
\begin{equation}
 \barT \tp^* \dfrac{d\tp}{dy}\Bigg|_{y_l}^{y_h} - \int_{y_l}^{y_h} dy \left( {\barT}\left| \dfrac{d\tp}{dy} \right|^2 + \left( \Ma^2 s^2 + \kz^2 \barT \right)\left|\tp\right|^2 \right) = 0.
\label{eq:span11_comp}
\end{equation} 
The pressure derivative can be written in terms of the normal velocity perturbation using the second expression in equation \ref{eq:span5_comp}, and the normal velocity perturbation at the wall is related to pressure through equation \ref{eq:span10_comp}. Therefore, equation \ref{eq:span11_comp} can be written as,
\begin{equation}
\begin{aligned}
 \sum_{y=y_l,y_h} &\dfrac{s^2 |\tp|^2}{\left(E_{eq} + D s + I s^2\right)} + \int_{y_l}^{y_h} dy \left( {\barT}\left| \dfrac{d\tp}{dy} \right|^2 + \left( \Ma^2 s^2 + \kz^2 \barT \right)\left|\tp\right|^2 \right) = 0,
\end{aligned}
\label{eq:span12_comp}
\end{equation}
where, $E_{eq} = E + T \kzsq + B \kz^4$ is positive. The imaginary part of the above equation can be simplified,
\begin{equation}
  \sum_{y=y_l,y_h} \dfrac{ \left( 2 s_R s_I E_{eq} + |s|^2 s_I D \right)|\tp|^2 }{ \left|E_{eq} + D s + I s^2\right|^2} + 2 s_R s_I \Ma^2 \int_{y_l}^{y_h} dy |\tp|^2 = 0 ,
\label{eq:span13_comp}
\end{equation}
where $s_R$ and $s_I$ are the real and imaginary parts of $s$, respectively. It is clear from the above equation that a neutral mode, i.e. $s_R = 0$, cannot exist if $D \neq 0$. For a non-dissipative wall ($D = 0$), both terms of equation \ref{eq:span13_comp} are positive, therefore their sum can be zero only if $s_R s_I = 0$. But all terms in the left hand side of equation \ref{eq:span12_comp} for $D = 0$, will become positive if $s_R \neq 0$ and $s_I = 0$, which is not possible. This leads to the following result,
 \begin{theoremcompliant}
 A purely span-wise mode is neutrally stable ($s_R = 0$) in the inviscid limit for compressible flows past non-dissipative ($D = 0$) compliant walls.
 \label{theorem:proposition6_compliant}
 \end{theoremcompliant}
Note that although in deriving the above result, the boundary terms are evaluated assuming that both walls are compliant, the proposition  will hold even for internal flows with one rigid wall or external flows. For an internal flow, if one wall is rigid the normal velocity is zero and by equation \ref{eq:span5_comp}, the derivative of pressure is zero, and therefore its contribution to the boundary term in equation \ref{eq:span11_comp} is zero. The same happens for external flows when $s_R \neq 0$, or $s_R = 0$ and $\Ma^2 s_I^2 < \kzsq \barT$ at the freestream. For an external flow with $s_R = 0$ and $\Ma^2 s_I^2 > \kzsq \barT$, the pressure perturbations do not vanish at infinity (see appendix \ref{subsecapp:bc_span}). However, evaluating the boundary terms in equation \ref{eq:span11_comp} at infinity (see equation \ref{eqapp:eq66} in appendix \ref{subsecapp:bc_span}), it can be noted that for $D = 0$, the imaginary part is identically zero and hence no additional contribution comes in equation \ref{eq:span13_comp} from the freestream, implying proposition \ref{theorem:proposition6_compliant} still holds. 
 
For dissipative compliant walls, i.e., $D > 0$,  we can rearrange the terms in equation \ref{eq:span13_comp} as,
\begin{equation}
\begin{aligned}
2 s_R s_I &\left( \sum_{y=y_l,y_h} \dfrac{ E_{eq}|\tp|^2 }{ \left|E_{eq} + D s + I s^2\right|^2} + \Ma^2 \int_{y_l}^{y_h} dy |\tp|^2 \right) \\
&\hspace{3cm}+ \left( \sum_{y=y_l,y_h} \dfrac{  |s|^2 |\tp|^2 }{ \left|E_{eq} + D s + I s^2\right|^2}\right) s_I D = 0 ,
\end{aligned}
\label{eq:span14_comp}
\end{equation}
where both terms inside the parentheses are positive. This implies that for $D > 0$, the equation can only be satisfied if $s_I = 0$. From equation \ref{eq:span12_comp}, if $s_I = 0$ then $s_R < 0$, since every other term is positive except the term multiplying $D$. 
This leads to the following result,
 \begin{theoremcompliant}
 Purely span-wise modes in compressible flows past dissipative ($D > 0$) compliant walls are never unstable in the inviscid limit.
 \label{theorem:proposition6b_compliant}
 \end{theoremcompliant} 
The above proposition holds for both internal flows with one rigid wall and external flows when $s_R \neq 0$. However, an $s_R = 0$ solution may be allowed with a dissipative wall if $\Ma^2 s_I^2 > \kzsq \barT$. In that case, an additional contribution appears in equation \ref{eq:span14_comp} that is proportional to the difference of the relative energies of the incoming and outgoing waves (see equation \ref{eqapp:eq65} in appendix \ref{subsecapp:bc_span}). Therefore, the flow can then sustain a neutral mode ($s_R = 0$) with a dissipative wall such that the energy dissipated at the wall is carried into the flow from the freestream, however, an unstable mode is not possible to be sustained.  

\section{Summary}
\label{sec:summary_compliant}

The general theorems of stability for compressible flows past compliant surfaces have been derived in this study. These theorems provide essential first-hand insights on how the stability characteristics modify due to flow compressibility and wall compliance. 
The theorems derived have some similarities and differences with the previously known results for incompressible compliant wall flows and compressible rigid wall flows. In this section, we summarize these and their consequences on the stability characteristics in the inviscid limit.

It is well known that a compressible flow past an object can sustain neutral modes with non-vanishing eigenfunctions far from the object provided they are supersonic in the freestream. This is different from the incompressible counterpart or for flows in rigid wall-bounded domains, where neutral modes have vanishing normal velocity at all boundaries. Physically, a non-vanishing normal velocity at infinity leads to a possibility that the energy generated due to the Reynolds stress discontinuity across a critical point in the domain is carried away into the freestream or vice-versa, thereby maintaining neutrality of the mode. Therefore, the compressible flow equivalent of the base velocity curvature, the quantity $d/dy((1/\barT) d\baru/dy)$, is not required to be zero at the critical point for neutral supersonic modes in an external flow, unlike subsonic neutral modes or all neutral modes in an internal flow (proposition 1). The presence of a compliant wall brings in an additional mechanism of energy dissipation, only if the wall is dissipative. Therefore, for neutral modes in an internal flow or subsonic neutral modes in an external flow (both of which having no additional energy exchange mechanisms at the non-compliant boundary), the quantity $\baru \:d/dy((1/\barT) d\baru/dy)$ is negative at the critical point if at least one compliant wall is dissipative (proposition 2). However, for neutral supersonic modes, the result cannot be generalised as the neutrality of the mode is maintained by a balance between the energy generated in the flow, the energy exchange at infinity and the energy dissipated at the wall. These results are qualitatively similar to incompressible flows past compliant walls, where the quantity $\baru \:d^2\baru/dy^2$ is required to be negative, for a dissipative wall, or zero, for a non-dissipative wall, at the critical point (see \cite{ref-kumaran-2021}) for the existence of neutral modes. 

In the absence of a critical point in the flow, there are no energy generation mechanisms within the flow. Therefore, a neutral mode can possibly exist only in an external flow where the energy dissipated at the compliant wall could be supplied by the neutral waves at infinity, provided they are supersonic at the freestream (proposition 3). This result does not have a counterpart in the incompressible limit, as neutral modes with non-vanishing eigenfunctions can exist only in compressible unbounded flows. For all other cases, finite wall dissipation expectedly produces stable modes when the real part of the wave-speed is outside the range of minimum and maximum of base velocity. 
Unstable modes can exist in compressible compliant wall flows if $\cR$ is in the range of $(\baru_L,\baru_U)$, where $\baru_L = \mbox{Min}(\mbox{Min}(\baru),0)$ and $\baru_U = \mbox{Max}(\mbox{Max}(\baru),0)$ (proposition 5). This result is same for incompressible compliant wall flows but is different from rigid wall compressible flows where the range is $(\minbaru,\maxbaru)$ instead. The equivalent of the Howard semi-circle theorem holds where the unstable eigenvalues are bound by a semi-circle in the complex $c$-space with the center at $(\baru_U + \baru_L)/2$ and radius $(\baru_U - \baru_L)/2$ (proposition 8).

Compressible shear flows generally have an infinite sequence of modes with $\cR$ outside the range of minimum and maximum of base velocity. In the presence of non-dissipative compliant walls, these modes are constrained to be neutral in the inviscid limit. For such modes, if the magnitude of their wave speed is greater than the free wave-speed of the compliant wall, it is required that $\chi = \barT( 1- \Ma_r^2)$ is negative somewhere in the domain (proposition 4). As wall compliance decreases to zero, the free wave-speed asymptotically approaches infinity. Therefore, in the limit of zero compliance, the above result does not seem to hold. However, this is not true because the same criteria can be derived as an existence condition for all neutral modes in a compressible rigid wall flow when $\cR$ is outside the range of $(\minbaru,\maxbaru)$ (see \cite{ref-deka-2022,ref-deka-2023}). For rigid wall compressible flows, the $\chi$-criteria can be used to compute critical Mach numbers below which all compressible modes are stable in the inviscid limit (see \cite{ref-deka-2022,ref-deka-2023}). However, for compressible flows past compliant walls, due to the added constraint on $\cR$, the theorem does not provide the same critical Mach numbers as those for the rigid wall case. Whether the critical Mach numbers for compliant wall flows will be higher or lower than those of the rigid wall modes can only be determined by numerical calculations. 

Additional results for purely spanwise modes can be obtained for compressible flows past compliant surfaces. For spanwise modes, the eigenfunctions are regular everywhere in the domain and so there is no energy generation mechanism within the flow. Therefore, in the presence of non-dissipative compliant walls, the modes are neutral just as in the case for rigid walls (proposition 9). In the presence of a dissipative compliant wall, all modes are stable, but there could exist a neutral mode in an external flow that is supersonic at the freestream if the energy dissipated at the surface is supplied by the freestream neutral waves (proposition 10).
 
\backmatter












\begin{appendices}
   
\section{Normal mode equations}
\label{secapp:normalmodeeqs_inv}

The linearised mass, $x$-momentum, $y$-momentum, $z$-momentum, and temperature equations for normal modes imposed on an inviscid compressible plane parallel flow are, 
\begin{equation}
\label{rho_mode_inv}
 \imath \kx ( \baru - c ) \trho + \frac{d \barrho}{dy} \tv + \barrho \left( \imath \kx \tu + \frac{d \tv}{dy} + \imath \kz \tw \right) = 0,
\end{equation}
\begin{equation}
\barrho \left(\imath \kx ( \baru - c ) \tu + \frac{d\baru}{dy} \tv \right) = -\frac{\imath \kx \tp}{\gamma \Ma^{2}},
\label{u_mode_inv}
\end{equation}
\begin{equation}
\barrho \imath \kx ( \baru - c ) \tv = -\frac{1}{\gamma \Ma^{2}} \frac{d\tp}{dy} ,
\label{v_mode_inv}
\end{equation}
\begin{equation}
\barrho \imath \kx ( \baru - c ) \tw = -\frac{\imath \kz \tp}{\gamma \Ma^{2}},
\label{w_mode_inv}
\end{equation}
\begin{equation}
\barrho \left(\imath \kx ( \baru - c ) \tT + \frac{d\barT}{dy} \tv \right) = - (\gamma - 1) \barrho \barT 
\left( \imath \kx \tu + \frac{d\tv}{dy} + \imath \kz \tw\right) .
\label{T_mode_inv}
\end{equation}
The equation of state is,
\begin{equation}
\label{p_mode_inv}
    \tp = \barrho \tT + \barT \trho.
\end{equation}

\section{Boundary conditions for the compressible Rayleigh equation}
\label{secapp:bc_rayleigh}

To solve the compressible Rayleigh equation, equation \ref{eq:eq35_inv-comp}, the boundary conditions for normal velocity is required at the boundaries. For rigid walls, the no-penetration condition implies that the normal velocity perturbations are zero. For unbounded flows, the velocity perturbations are in general required to be bounded. In order to determine the boundary conditions for external flows, we solve equation \ref{eq:eq35_inv-comp} for a uniform flow conditions, i.e, $\baru = U_\infty, \barT = T_\infty$. The solution (see \cite{ref-mack-87}) can be expressed as,
\begin{equation}
 \tv = A \exp{ \left(\sqrt{(\kxsq + \kzsq)( 1 - \Ma_{r,\infty}^2)}y\right)} + B \exp{ \left(- \sqrt{(\kxsq + \kzsq)( 1 - \Ma_{r,\infty}^2)}y\right)},
\label{eq:ray_sol_freestream}
\end{equation}  
where, $A$ and $B$ are constants determined from the boundary conditions and $\Ma_{r,\infty}$ is the relative Mach number defined in equation \ref{eq:def_ma_r}, evaluated at the free-stream conditions. From the form of the solution, it can be easily inferred that as long as the arguments of the exponential functions have a real part, the exponentially decaying solution is the only one that remains bounded and hence allowed, implying that in the limit $y \rightarrow \infty$, the zero normal velocity boundary condition applies. This is the case for any non-neutral mode ($\cI \neq 0$) and for neutral modes ($\cI = 0$) that are subsonic, i.e., $\Ma_{r,\infty} < 1$. However, for supersonic neutral modes, $\Ma_{r,\infty} > 1$, the solution becomes an oscillatory function of $y$ and therefore, does not decay to zero as $y \rightarrow \infty$.   

For the results derived in section \ref{sec:inviscidflow_compliant}, the term $(c\tv^*/\barchi) d\tv/dy$ (equation \ref{eq:eq42a_inv-comp}) is required to be evaluated at the boundary, where $^*$ denotes complex conjugate. For non-neutral or subsonic neutral modes, the term vanishes in the limit $y \rightarrow \infty$. For supersonic neutral modes, substituting the solution from equation \ref{eq:ray_sol_freestream}, we get,
\begin{equation}
\begin{aligned}
 \frac{c\tv^*}{\barchi} \frac{d\tv}{dy} = -\cR\sqrt{\frac{(\kxsq + \kzsq)}{(\Ma_{r,\infty}^2 - 1)}}& \Bigg(\imath(|A|^2 - |B|^2)\\
	& - \mbox{Im} \left(2A B^* \exp(2\imath \sqrt{(\kxsq + \kzsq)( 1 - \Ma_{r,\infty}^2)} y) \right) \Bigg),
\end{aligned}
\label{eqapp:eq61}
\end{equation}
where $\mbox{Im}(\cdot)$ denotes the imaginary part, and the relation $\barchi = \barT(1 - \Ma_r^2)$ is used. It is clear that the imaginary part of $(c\tv^*/\barchi) d\tv/dy$ is non-zero. 

We also evaluate the boundary term on the left hand side of equation \ref{eq:eq37_inv-comp} expressed in terms of the function $\tg= \tv/(\baru - c)$. At the freestream, $d\baru/dy = 0$, therefore we can write,
\begin{equation}
 \frac{(\baru - c)^2 \tga}{\chi}\frac{d\tg}{dy} = \frac{\tva}{\chi}\frac{d\tv}{dy},
\label{eqapp:eq62}
\end{equation}
which from equation \ref{eqapp:eq61} has a non-zero imaginary part. Therefore, considering the imaginary part of equation \ref{eq:eq64_inv-comp} for an external flow past a dissipative compliant walls ($D = 0$), the additional contribution to the boundary term of equation \ref{eq:eq37_inv-comp} comes for a supersonic neutral mode from the imaginary part of equation \ref{eqapp:eq62}, i.e.,
\begin{equation}
 -\frac{\cR D}{\kx} - \sqrt{\frac{(\kxsq + \kzsq)}{(\Ma_{r,\infty}^2 - 1)}} (|A|^2 - |B|^2) = 0.
\label{eqapp:eq63}
\end{equation}

\subsection{Purely spanwise modes}
\label{subsecapp:bc_span}
For purely spanwise perturbations, the normal modes form is shown in equations \ref{eq:span1_comp}-\ref{eq:span4_comp}, and an equation for the pressure perturbation is obtained in the inviscid limit, shown in equation \ref{eq:span7_comp}. At rigid walls, the no penetration condition reduces to the requirement that the derivative of pressure vanishes as evident from the relation between normal velocity and pressure in equation \ref{eq:span5_comp}. For an external flow, equation \ref{eq:span7_comp} can be solved in the freestream to obtain the solution,
\begin{equation}
\tp = C \exp{ \left(\sqrt{\kzsq + \frac{\Ma^2 s^2}{\barT}}\: y\right)} + D \exp{ \left(-\sqrt{\kzsq + \frac{\Ma^2 s^2}{\barT}}\: y\right)},
\label{eq:span_sol_freestream}
\end{equation}
where, $C$ and $D$ are constants determined from the boundary conditions. If we write, $s = s_R + \imath s_I$, then it is clear from equation \ref{eq:span_sol_freestream} that if $s_R \neq 0$, an exponentially decaying solution is allowed, for which $\tp = 0$ in the limit $y\rightarrow\infty$. For $s_R = 0$, and $\Ma^2 s_I^2 > \kzsq \barT$, an oscillatory solution is obtained, which does not decay to zero as $y\rightarrow\infty$. In that case, the boundary term in equation \ref{eq:span11_comp} obtained after substituting the freestream solution \ref{eq:span_sol_freestream} becomes,
\begin{equation}
\barT \tp^* \frac{d\tp}{dy} = (\Ma^2 s_I^2 - \kzsq \barT) \left( \imath(|A|^2 - |B|^2) - \mbox{Im} \left(2A B^* \exp \left(2\imath \sqrt{\frac{\Ma^2 s_I^2}{\barT} - \kzsq} y\right) \right) \right)
\label{eqapp:eq65}
\end{equation}
The above term would appear in the left hand side of equation \ref{eq:span12_comp} for an external flow. For a non-dissipative wall ($D = 0$), equation \ref{eq:span12_comp} has no imaginary part, which implies that $|A| = |B|$. This is expected as for a neutral wave in the absence of an source of energy (like across a critical point for streamwise perturbations), the incoming and outgoing waves have to carry equal energies. Therefore, for $s_R = 0$, and $\Ma^2 s_I^2 > \kzsq \barT$, equation \ref{eqapp:eq65} simplifies for a non-dissipative compliant wall ($D = 0$) to,
\begin{equation}
\barT \tp^* \frac{d\tp}{dy} = - (\Ma^2 s_I^2 - \kzsq \barT) \mbox{Im} \left(2|A|^2 \exp \left(2\imath \sqrt{\frac{\Ma^2 s_I^2}{\barT} - \kzsq} y\right) \right) .
\label{eqapp:eq66}
\end{equation}

\end{appendices}

\bibliography{ref}

\end{document}